\documentclass[iop,twocolumn]{emulateapj}

\newcounter{species}
\def\ion#1#2{\setcounter{species}{#2}#1$\;${\scriptsize\Roman{species}}\relax}

\usepackage{xspace}

\newcommand{\naa}{\ion{Na}{1}\xspace}
\newcommand{\ca}{\ion{Ca}{2}\xspace}
\newcommand{\ewna}{$\mathrm{EW}_{\mathrm{Na}}$\xspace}
\newcommand{\ewca}{$\mathrm{EW}_{\mathrm{Ca}}$\xspace}
\newcommand{\teff}{$\mathrm{T}_{\mathrm{eff}}$\xspace}
\newcommand{\mk}{$M_K$\xspace}
\newcommand{\dewna}{$\Delta\mathrm{EW}_{\mathrm{Na}}$\xspace}
\newcommand{\dewca}{$\Delta\mathrm{EW}_{\mathrm{Ca}}$\xspace}
\newcommand{\dmk}{$\Delta M_K$\xspace}

\usepackage{color}
\usepackage{graphicx}

\shorttitle{M dwarf parameters from NIR spectra}
\shortauthors{Ryan C Terrien et al.}
\begin{document} 
\title{M dwarf luminosity, radius, and $\alpha$-enrichment from $I$-band spectral features}

\author{Ryan C. Terrien\altaffilmark{1,2,3},
Suvrath Mahadevan\altaffilmark{1,2,3}, 
Chad F. Bender\altaffilmark{1,2},
Rohit Deshpande\altaffilmark{1,2}, 
\& Paul Robertson\altaffilmark{1,2}
}

\slugcomment{Accepted to ApJL, March 2015}

\email{rct151@psu.edu}
\altaffiltext{1}{Department of Astronomy and Astrophysics, The Pennsylvania State University, 525 Davey Laboratory, University Park, PA 16802, USA.}
\altaffiltext{2}{Center for Exoplanets and Habitable Worlds, The Pennsylvania State University, University Park, PA 16802, USA.}
\altaffiltext{3}{The Penn State Astrobiology Research Center, The Pennsylvania State University, University Park, PA 16802, USA.}

\begin{abstract}
Despite the ubiquity of M dwarfs and their growing importance to studies of exoplanets, Galactic evolution, and stellar structure, methods for precisely measuring their fundamental stellar properties remain elusive. Existing techniques for measuring M dwarf luminosity, mass, radius, or composition are calibrated over a limited range of stellar parameters or require expensive observations. We find a strong correlation between the $K_S$-band luminosity ($M_K$), the observed strength of the $I$-band sodium doublet absorption feature, and [Fe/H] in M dwarfs without strong H$\alpha$ emission. We show that the strength of this feature, coupled with [Fe/H] and spectral type, can be used to derive M dwarf $M_K$ and radius without requiring parallax. Additionally, we find promising evidence that the strengths of the $I$-band sodium doublet and the nearby $I$-band calcium triplet may jointly indicate $\alpha$-element enrichment. The use of these $I$-band features requires only moderate-resolution near-infrared spectroscopy to provide valuable information about the potential habitability of exoplanets around M dwarfs, and surface gravity and distance for M dwarfs throughout the Galaxy. This technique has immediate applicability for both target selection and candidate planet host system characterization for exoplanet missions such as \textit{TESS} and \textit{K2}.

\end{abstract}


\maketitle

\section{Introduction}
\label{sec:intro}
M dwarfs account for the majority of nearby stars \citep[e.g.][]{2002AJ....124.2721R}, but due to their relative faintness and complex molecular spectra, precise ($<10\%$) measurements of their stellar parameters remain challenging. Measurements of the stellar mass, radius, luminosity, and composition of M dwarfs are important for studies of stellar structure \citep{Torres:2009eo}, Galactic evolution \citep{Bochanski:2013ky}, and particularly for studies of exoplanets, since indirectly derived planetary characteristics and histories depend on accurate knowledge of the host star. M dwarfs are attractive planet search targets because their smaller sizes and lower luminosities result in larger radial velocity (RV) and transit depth signatures compared to Sun-like stars. They appear to host numerous planetary systems \citep{Dressing:2015wy} and are prioritized in planned exoplanet surveys \citep[e.g.][]{Ricker:2015ie}. However, the characterization of the resulting planets and whether they are in the habitable zone \citep{Kasting:1993hw} could be clouded by lack of efficient and precise stellar characterization. 

The complexity of M dwarf spectra encumbers spectroscopic modeling and the measurement of stellar parameters. Numerous techniques have been developed to circumvent this difficulty, by relating M dwarf parameters to easily-measured photometry or moderate-resolution spectroscopic features. These relations are calibrated using stars for which the parameters are known a-priori, such as M dwarfs with interferometrically-measured radii \citep{Boyajian:2012eu,Mann:2013fv,Newton:2014uy} or M dwarf companions to higher-mass stars with well-measured compositions \citep[e.g.][]{2005A&A...442..635B,2009ApJ...699..933J,RojasAyala:2012fb,Mann:2013kj,2014AJ....147...20N}. 

Determinations of M dwarf radius, luminosity, and mass can be made through empirical relations between these parameters and M dwarf effective temperature (T$_{\mathrm{eff}}$), which can be estimated from moderate-resolution spectral features \citep{Boyajian:2012eu,Mann:2013fv}. However, this technique is limited by systematic errors when using stellar models to predict M dwarf parameters from T$_{\mathrm{eff}}$ and [Fe/H] \citep[e.g.][]{Boyajian:2012eu}, by the paucity of interferometric calibrator stars, and by a large range of M dwarf parameters at a given estimated \teff \citep[e.g.][]{Newton:2014uy}. Combining improved \teff measures and [Fe/H] enables better M dwarf parameters, but requires broadband flux-calibrated spectra \citep{Mann:2015us}.

One can also estimate M dwarf parameters from $K_s$-band luminosity ($M_K$), which is well-correlated with stellar mass and insensitive to metallicity \citep{Delfosse:2000ur}. This allows mass measurements of nearby M dwarfs with parallaxes, but parallaxes for many nearby M dwarfs may not be available for several years \citep{Perryman:2001cp}. For M dwarfs without parallax measurements, some success has been achieved in determining $M_K$ using high-resolution spectra \citep{Pineda:2013wy}, but such spectra are observationally expensive. Methods that do not require parallax, high-resolution spectra, or absolute flux measurements of the target stars, and which can be applied immediately are highly desirable.

We have developed a simple, powerful technique, based on the strengths of the $I$-band neutral sodium doublet ($\sim820$~nm, hereafter \naa) and singly-ionized calcium triplet ($\sim860$~nm, hereafter \ca), to enable determination of M dwarf $M_K$ and radius, and convey information about M dwarf surface gravity and possibly $\alpha$-enrichment. We present here an observational study that shows the potential applicability of these $I$-band features in conjunction with measurements of M dwarf \teff and [Fe/H], all obtained from moderate-resolution near-infrared spectra.

\section{Observations and Methods}
\label{sec:obs}
The 342 stars used here were observed in our IRTF-SpeX \citep{2003PASP..115..362R} M dwarf survey. \citet{2012ApJ...747L..38T} detailed the observations and data reduction. We also include spectra of nine nearby M dwarfs from \citet{Mann:2013fv}. All spectra have $R\sim2000$ and span $0.8-2.4\mu$m. We measured [Fe/H] for each star using the weighted (by signal-to-noise, S/N) average of the $J,H,K_s$ NIR spectral calibrations of \citet{Mann:2013kj} (for M5 and earlier) and \citet{Mann:2014bz} (later than M5), which have $<0.1$~dex precision. We also measured the \teff-sensitive H$_2$O-K2 index and associated NIR M spectral type (NIR SpT) explored in \citet{RojasAyala:2012fb} and \citet{2014AJ....147...20N}.

For each target, we compiled parallax \citep[from the RECONS\footnote{http://www.recons.org} database,][]{vanAltena:1995ts,Gould:2004jo,vanLeeuwen:2007dc,Dittmann:2014cr,Gatewood:2008hu,vonBraun:2011ke,AngladaEscude:2012gk}, proper motion \citep{Zacharias:2013cf} and RV \citep{Chubak:2012tv,2014AJ....147...20N} where available. We used H$\alpha$ equivalent width measurements \citep[e.g.][]{Gizis:2002ej,Riaz:2006du,Gaidos:2014if} to constrain activity levels for our M dwarfs.

\begin{figure*}
\begin{center}
\includegraphics[width=.7\paperwidth]{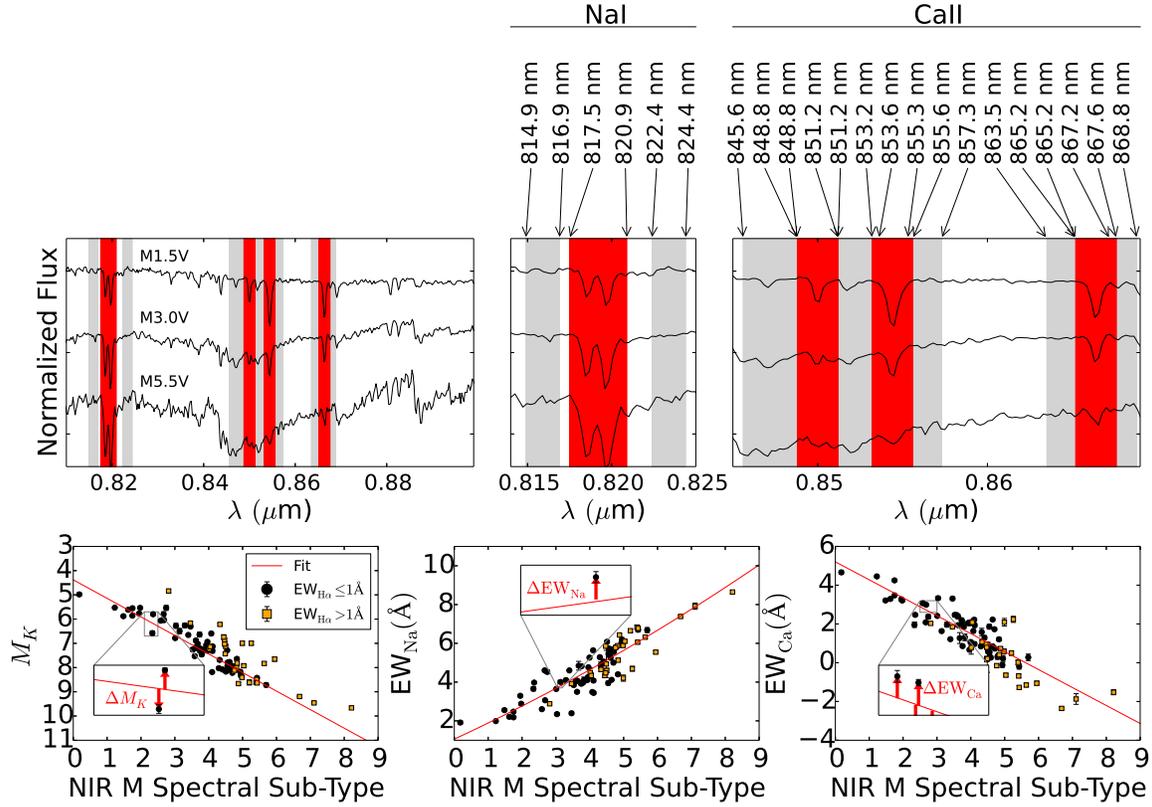}
\caption{\label{fig:features}
\textit{Top:} The feature (red) and continuum (gray) definitions for the quantities used in this work. \textit{Bottom:} The strengths of the features with NIR M subtype (a proxy for \teff). For \naa, \ca, and \mk, we fit a low-order polynomial to these quantities for a subset of M dwarfs in our sample, and defined \dewna, \dewca, \dmk as the departure from this fit (see Section \ref{sec:obs}).}
\end{center}
\end{figure*}

Our technique relies on measurements of the pseudo-equivalent widths (EWs) of the \naa and \ca features. The \naa doublet is a known gravity indicator for M dwarfs \citep{Schiavon:1997uo,Martin:2010cx,Schlieder:2012iw} and both the \naa doublet and the \ca triplet have shown possible sensitivity to H$\alpha$-based activity levels \citep{Kafka:2006kw}. The definitions of the features and continuum regions are shown in Figure \ref{fig:features}. For the \naa doublet and each component of the \ca triplet, we fit a line to the two neighboring continuum regions (Figure \ref{fig:features}), and calculated the EW of the feature relative to this line. Our final \ewca value is the sum of the three components. We define positive EW as absorption for the \ca and \naa features. The measured and literature parameters for each star are listed in Table \ref{tab:data1}. For six M dwarfs observed in our program and \citet{Mann:2013fv}, we find no offset for the \naa feature, but that our measurements of \ewca are systematically $0.96$~\AA~higher, possibly due to our use of SpeX ($R\sim 2000$) in the $I$-band instead of SNIFS \citep[][$R\sim 1000$]{Lantz:2004dk} used in \citet{Mann:2013fv}. We apply this correction for the 9 stars from \citet{Mann:2013fv} that are not in our sample.

\begin{deluxetable*}{lcccccccc}
\tabletypesize{\tiny}
\tablecaption{M Dwarf Parameters and Measurements \tablenotemark{a}\label{tab:data1}}
\tablehead{\colhead{2MASS ID} & 
\colhead{[Fe/H]} & 
\colhead{NIR SpT} & 
\colhead{EW$_{\mathrm{Na}}$} & 
\colhead{EW$_{\mathrm{Ca}}$} & 
\colhead{$R_{\mathrm{interfer.}}$\tablenotemark{b}} & 
\colhead{$R_{\mathrm{\pi}}$\tablenotemark{c}} & 
\colhead{$R_{\mathrm{Na}}$\tablenotemark{d}} & 
\colhead{M13 Flag\tablenotemark{e}} \\ 
& & & (\AA) & (\AA) & ($R_{\odot}$) & ($R_{\odot}$) & ($R_{\odot}$) &  }  
\startdata
00064325$-$0732147 &  $ -0.11 $  &  $ 5.99$  &  $ 7.46 \pm 0.05 $  &  $ 0.24 \pm 0.07 $  & \ldots & \ldots & \ldots & \ldots \\ 
00085391$+$2050252 &  $ 0.00 $  &  $ 4.93$  &  $ 5.38 \pm 0.11 $  &  $ 0.83 \pm 0.15 $  & \ldots & \ldots & \ldots & \ldots \\ 
00115302$+$2259047 &  $ 0.15 $  &  $ 3.62$  &  $ 4.08 \pm 0.09 $  &  $ 2.38 \pm 0.11 $  & \ldots & \ldots & \ldots & \ldots \\ 
00165629$+$0507261 &  $ -0.14 $  &  $ 4.52$  &  $ 3.90 \pm 0.09 $  &  $ 1.80 \pm 0.11 $  & \ldots & \ldots & \ldots & \ldots \\ 
00182256$+$4401222 &  $ -0.26 $  &  $ 1.89$  &  $ 2.56 \pm 0.10 $  &  $ 4.45 \pm 0.13 $  &  $ 0.386 \pm 0.002 $  &  $ 0.395 \pm 0.004 $  &  $ 0.416 \pm 0.027 $  & \ldots \\ 
00182549$+$4401376 &  $ -0.09 $  &  $ 4.10$  &  $ 5.64 \pm 0.05 $  &  $ 2.22 \pm 0.08 $  & \ldots & \ldots & \ldots & \ldots \\ 
00283948$-$0639481 &  $ -0.11 $  &  $ 4.18$  &  $ 3.25 \pm 0.14 $  &  $ 2.01 \pm 0.15 $  & \ldots & \ldots & \ldots & \ldots \\ 
00313539$-$0552115 &  $ -0.12 $  &  $ 3.68$  &  $ 4.06 \pm 0.09 $  &  $ 2.12 \pm 0.13 $  & \ldots & \ldots & \ldots & \ldots \\ 
00321574$+$5429027 &  $ -0.05 $  &  $ 4.49$  &  $ 3.85 \pm 0.07 $  &  $ 1.55 \pm 0.10 $  & \ldots & \ldots & \ldots & \ldots \\ 
00383388$+$5127579 &  $ -0.24 $  &  $ 2.83$  &  $ 3.30 \pm 0.07 $  &  $ 3.25 \pm 0.09 $  & \ldots & \ldots & \ldots & \ldots \\ 
\multicolumn{9}{c}{\ldots}
\enddata
\tablenotetext{a}{Table \ref{tab:data1} is published in its entirety (including parameters from the literature) in the electronic edition of \textit{ApJ Letters}, a portion is shown here for guidance regarding its form and content.}
\tablenotetext{b}{Interferometric radius from \citet{Boyajian:2012eu}.}
\tablenotetext{c}{Interpolated at parallax-based $M_K$ and 5~Gyr age in Dartmouth model grid.}
\tablenotetext{d}{Interpolated at predicted $M_K$ (see Section \ref{sec:mk}) in Dartmouth model grid.}
\tablenotetext{e}{Spectrum from \citet{Mann:2013fv}}.
\end{deluxetable*} 

To explore the empirical behavior of \mk, \ewna, and \ewca as a function of NIR SpT, we used a subset of 84 M dwarfs with high S/N spectra, parallax measurements with precisions better than 5\%, no indications of multiplicity, and either low (EW$_{\mathrm{H}\alpha} < $1~\AA,~$n=57/84$) or high (EW$_{\mathrm{H}\alpha} \geq $1~\AA,~$n=27/84$) activity levels as evidenced by the strength of H$\alpha$ emission \citep[e.g.][]{West:2004kd}. We fit each of these quantities with a low-order polynomial as shown in Figure \ref{fig:features}, and defined \dmk, \dewna, and \dewca as the departure from these fits at a given NIR SpT. For $M_K$ and \ca, we used only the low-activity subset ($n=57$), as these stars exhibit structural changes related to activity \citep[][and references therein]{Stassun:2012ft} and we observe the \ca feature in emission in some M dwarfs. We used all 84 stars for the fit to EW$_{\mathrm{Na}}$, because there is no clear evidence of activity measurably affecting the strength of the \naa feature in our data. The resulting definitions and low-order fits are:

\begin{eqnarray}
\Delta \mathrm{EW}_{\mathrm{Na}} & = & \mathrm{EW}_{\mathrm{Na}} - ( 1.07 + 0.78 \, \mathrm{[NIR \, SpT]} \\
& & \qquad {} + 0.03 \, \mathrm{[NIR \, SpT]}^2 ) \nonumber \\
\Delta \mathrm{EW}_{\mathrm{Ca}} & = & \mathrm{EW}_{\mathrm{Ca}} - ( 5.23 - 0.93 \,\mathrm{[NIR \, SpT]} ) \\
\Delta M_K & = & M_K - ( 4.36 - 0.77 \, \mathrm{[NIR \, SpT]} ).
\end{eqnarray}

\section{Results and Discussion}

\subsection{A relationship between $M_K$, \ewna, and [Fe/H]}
\label{sec:mk}
\begin{figure*}
\begin{center}
\centerline{\hspace{-1cm}\includegraphics[width=.7\paperwidth]{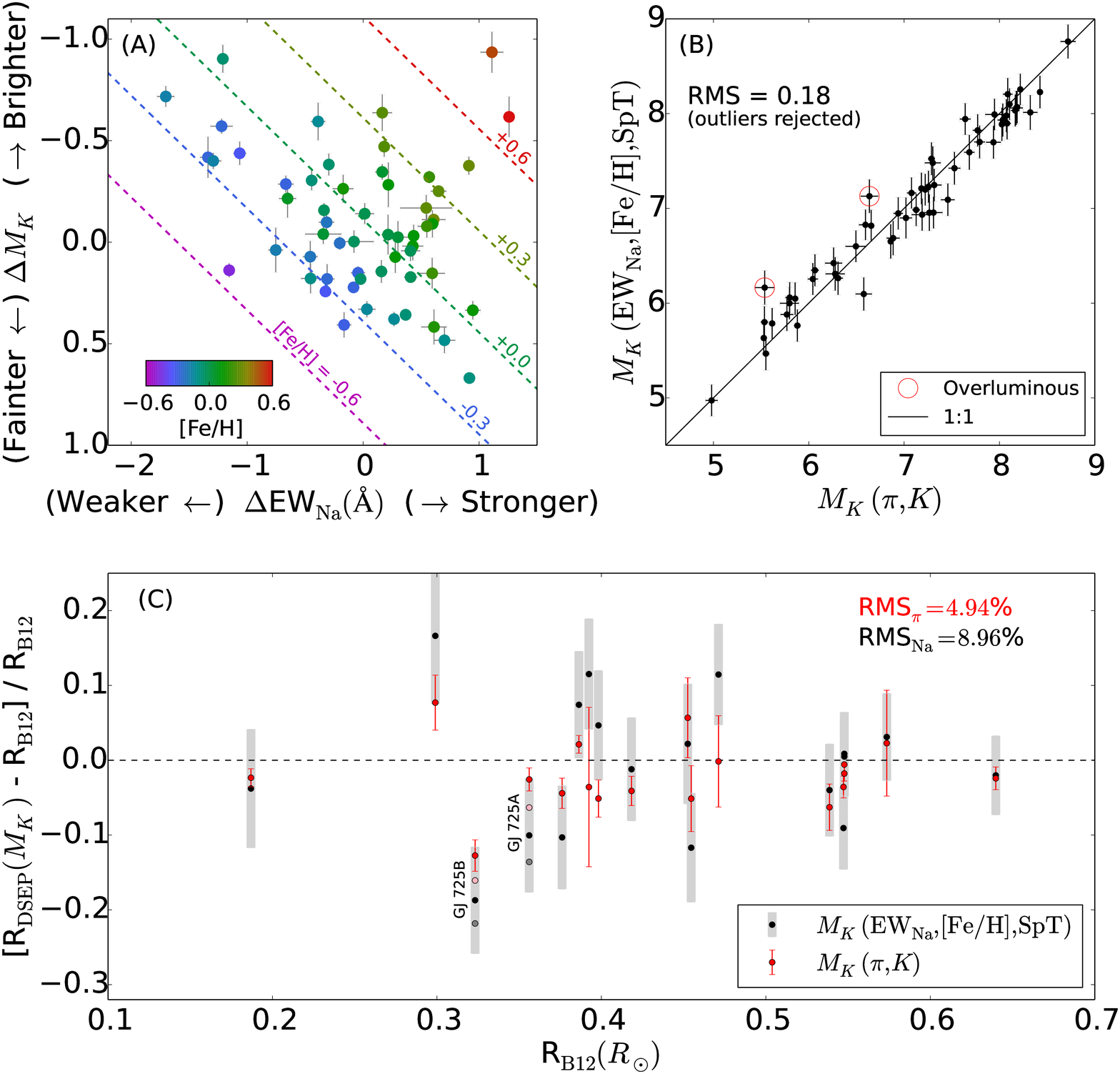}}
\caption{\label{fig:radlum}
	\textit{(A):} For 57 well-characterized nearby M dwarfs, the relationship between \dmk, \dewna, and [Fe/H], along with iso-[Fe/H] contours from our model that predicts \dmk from \dewna and [Fe/H]. \textit{(B):} For the same set of stars, the known parallax ($\pi$)-based $M_K$ and our predicted $M_K$ based on \ewna, [Fe/H] and NIR SpT. A 1:1 relation is shown; two overluminous targets that are likely binaries are highlighted. \textit{(C):} For 18 M dwarfs with interferometrically-measured radii \citep{Boyajian:2012eu}, the fractional difference between the calculated (R$_{\mathrm{DSEP}}$) and measured radii (R$_{\mathrm{B12}}$). To derive radius, we interpolate on the Dartmouth model grid at an $M_K$ and [Fe/H] appropriate to each star. We used two different values for $M_K$ for each star: the measured parallax ($\pi$)-based $M_K$ (black, gray error bars) and the $M_K$ predicted from our calibration of the \naa doublet (red). The $R$ values from predicted $M_K$ values have $\mathrm{RMS}_{\mathrm{Na}} \approx 9\%$, compared to $\mathrm{RMS}_{\pi} \approx 5\%$  from parallax-based $M_K$. We note that the mean error in R$_{\mathrm{B12}}$ is $\sim2\%$. Both sets of radii measurements are consistent with $<2\%$ offset from the Dartmouth models. GJ 725, a system with \ca and \naa feature strengths suggestive of $\alpha$-enrichment, is highlighted; the Dartmouth models with [$\alpha$/Fe]$=+0.2$ are shown in black and red while the [$\alpha$/Fe]$=0.0$ results are indicated in gray and pink (\citet{2008ApJS..178...89D} explains details regarding $\alpha$-enrichment in the Dartmouth models). For all other systems we used $[\alpha/\mathrm{Fe}]=0.0$.
}
\end{center}
\end{figure*}

We find that a first-order linear model based on $\Delta\mathrm{EW}_{\mathrm{Na}}$ and [Fe/H] yields an excellent predictor of \dmk (Figure~\ref{fig:radlum}A) in the low-activity subset. The resulting equation is:
\begin{equation}
\Delta M_K = -1.66 \, \mathrm{[Fe/H]} + 0.55 \, \Delta\mathrm{EW}_{\mathrm{Na}} - 0.11,
\end{equation} 
which has an adjusted squared correlation coefficient $R^2=0.64$ and residual scatter of 0.18~mag (excluding two overluminous stars that we suspect are binaries: 2MASS J04223199+1031188, 2MASS J02532611+1724324, Figure \ref{fig:radlum}B). This approximately halves the intrinsic $\sim0.35$~mag scatter in \mk from NIR SpT (Figure \ref{fig:features}). We explored alternative definitions of the \naa features \citep{Schlieder:2012iw,Martin:2010cx}, and the quality of the resulting fits differs negligibly from that reported here.

This model directly reveals that \ewna is sensitive to small changes in $M_K$ at a given NIR SpT. This is primarily due to the strong sensitivity of this feature to surface gravity at low \teff, a behavior that has been thoroughly explored in the context of youth indicators for low-mass stars \citep{Schlieder:2012iw}. With the inclusion of [Fe/H] in our model, we disentangle the gravity and [Fe/H] sensitivity of the \naa feature, which goes from being a binary indicator of youth or low gravity to having a much finer sensitivity to luminosity and surface gravity. Thus, this feature is a precise and easily-measured gravity indicator for M dwarfs, a stellar regime in which established precision gravity measurement techniques \citep[e.g.~flicker,][]{Bastien:2013ie} are not yet applicable.

Our interpretation that the \dmk-\dewna-[Fe/H] relationship results from the gravity-sensitivity of \ewna depends on the observed insensitivity to gravity of our [Fe/H] measurements. Although the \citet{Mann:2013kj} [Fe/H] calibrations employ gravity-sensitive features, we note that the \dmk-\dewna-[Fe/H] relationship is present (with similar coefficients) with different indicators of [Fe/H] (e.g.\ the individual $H$ and $K$-band calibrations), which are based on different features with varied behaviors with gravity. We also find this relationship using [Fe/H] derived from higher-mass companions, which are available for 21 of our stars from \citet{2005ApJS..159..141V}, and which are immune to effects within the M dwarf spectrum. We conclude that gravity-related systematic errors in our [Fe/H] measurements are insignificant.
	
We considered that our NIR SpT measurements could similarly have systematic errors that impact our interpretation of the \dmk-\dewna-[Fe/H] relationship. We note that our NIR SpTs are based on the H$_2$O-K2 index, which is known to correlate with \teff to $<80$K \citep[e.g.][]{Mann:2013fv}. By contrast, the \dmk-\dewna-[Fe/H] relationship extends over $\sim1.5$~mag in \dmk, which would equate to many hundreds of degrees in \teff. We conclude that, within their stated errors, our [Fe/H] and \teff indicators are accurately accounting for their contribution to the behavior of \ewna. Therefore, the well-established gravity-sensitivity of \ewna is the most reasonable explanation for its correlation with \dmk.

This model efficiently provides $M_K$ for M dwarfs without parallaxes, thereby enabling ``spectroscopic distance'' and fundamental parameter measurements of M dwarfs. As an illustration, we note that for one of the best-characterized M dwarfs in the \textit{Kepler} sample, Kepler~138 (KOI-314), we predict $M_K=5.42\pm0.18$ based on a single moderate-resolution NIR spectrum. This is in excellent agreement with the $M_K=5.39\pm0.25$ measured in \citet{Pineda:2013wy}, which is based on six high-resolution spectra of this star and library of more than 100 similar high-S/N spectra. Through the tight relations between $M_K$ and stellar mass \citep[e.g.][]{Delfosse:2000ur} and radius, the model we present can help to provide determinations of M dwarf parameters and the location of the habitable zone for M dwarf exoplanet hosts. Compared with existing techniques for measuring these parameters \citep{Mann:2013fv,Newton:2014uy}, our technique benefits from a large set of calibrators, and requires only a single, non-flux calibrated, moderate-resolution NIR spectrum.

\subsection{M dwarf radius from $M_K$}
\label{sec:radius}
After establishing that the \naa feature, [Fe/H], and NIR SpT can be used to determine M dwarf $M_K$, we explored extending the technique to measuring M dwarf radius, using 18 nearby M dwarfs with interferometrically measured radii (Figure~\ref{fig:radlum}C). We interpolated the Dartmouth Stellar Evolution Program \citep[DSEP,][]{2008ApJS..178...89D} models at the [Fe/H] and $M_K$ of each star to predict radius, using both the parallax-based $M_K$ and the spectroscopic $M_K$ indicated by the techniques in Section \ref{sec:mk}. We compared these radii with the interferometric radii \citep{Boyajian:2012eu}. We note that the mean error in the interferometrically measured radii themselves is approximately $2\%$. We found that we could derive radius to $\sim5$\% when using the parallax-based $M_K$, and to $\sim10$\% with just the information available from the NIR spectra. We find no evidence for a $>2\%$ systematic offset between $M_K$-based model-predicted and measured M dwarf radii in these 18 stars, in contrast to interpolations at L$_{\mathrm{Bol}}$ and \teff \citep[e.g.\ Section 5 of][]{Boyajian:2012eu}, demonstrating the advantages of using \mk and a possible path toward reconciling low-mass stellar models and observations. Using \mk to derive radius is therefore an accurate, precise, and efficient method that can be applied to M dwarfs well beyond the reach of current interferometric facilities.

\subsection{Measurements of M dwarf $\alpha$-enrichment?}
\label{sec:alpha}
\begin{figure*}
\begin{center}
\includegraphics[width=.7\paperwidth]{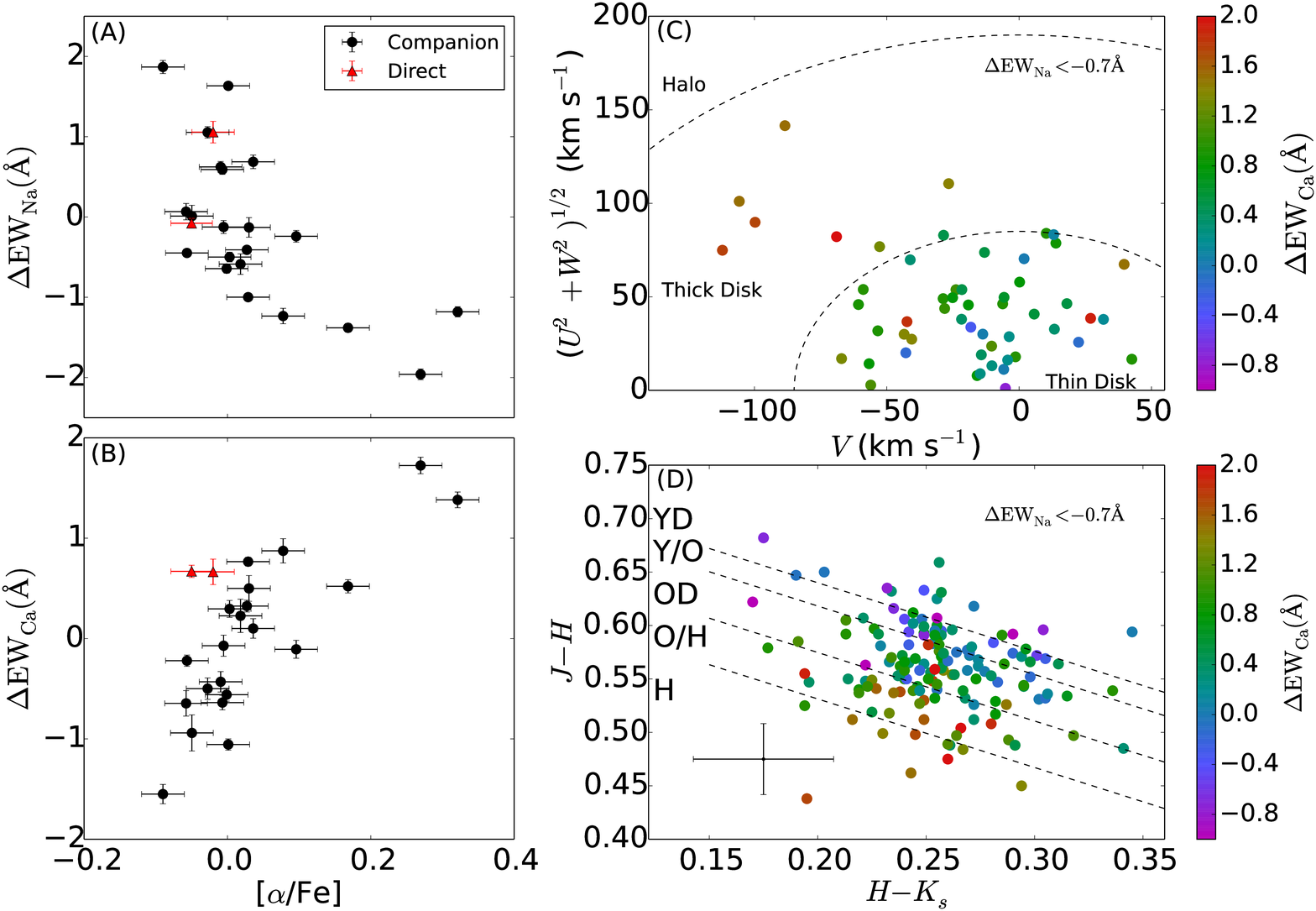}
\caption{\label{fig:alpha}
The behavior of the \ca and \naa features with $\alpha$-enrichment. \textit{(A,B):} For 23 M dwarfs, the reported $[\alpha/\mathrm{Fe}]$ parameter is the weighted averages of literature measurements of [Si/Fe] and [Ti/Fe] of a higher-mass common proper motion companion or the M dwarf itself, as a function of \dewna and \dewca. High \dewca and low \dewna are clearly associated with $\alpha$-enrichment. \textit{(C):} The Toomre diagram for M dwarfs with well-measured UVW velocities, and nominal Galactic population boundaries. Stars that are dynamically consistent with being old also have high \dewca. \textit{(D):} The color-color diagram with delineations of different populations \citep{1992ApJS...82..351L}. M dwarfs with high \dewca have colors consistent with old ages.
}
\end{center}
\end{figure*}

The $\alpha$-elements, such as silicon and titanium, can be formed in Type II supernovae, and they are more abundant (``enriched") relative to iron in older stellar populations. To explore the sensitivity of the \naa and \ca features to $\alpha$-enrichment, we first considered a subset of 23 M dwarfs from our sample with $\alpha$-enrichment ([Ti/Fe] and [Si/Fe]) measurements in the literature from either direct spectroscopy \citep{2005MNRAS.356..963W,Chavez:2009ih} or based on spectroscopy of higher-mass companions \citep{2005ApJS..159..141V}. We found that (Figure~\ref{fig:alpha}~A,B) $\alpha$-enriched stars have higher (stronger absorption) \dewca and lower (weaker absorption) \dewna, as expected since calcium is an $\alpha$-element and sodium is not. Moreover, for the subset of our M dwarfs where measurements are available, high \dewca and low \dewna are also consistent with kinematic and color-based indicators of old age. A Toomre diagram \citep[Figure~\ref{fig:alpha}~C,~][]{Sandage:1987el}, which shows stellar Galactocentric vertical ($U^2$), radial ($W^2$), and rotational ($V^2$) kinetic energies, can dynamically separate older and younger stars: older stars have a wide distribution of kinetic energies, while younger stars have lower kinetic energies. The lower metallicities of older stars also separate these stars well in the ($J-H,H-K$) color-color plane (Figure~\ref{fig:alpha}~D), due to H$^-$ continuum and H$_2$O opacity effects \citep{1992ApJS...82..351L}. These observational indications of abundance and age suggest that the \naa and \ca features jointly indicate $\alpha$-enrichment in M dwarfs.

\begin{figure*}
\begin{center}
\includegraphics[width=.7\paperwidth]{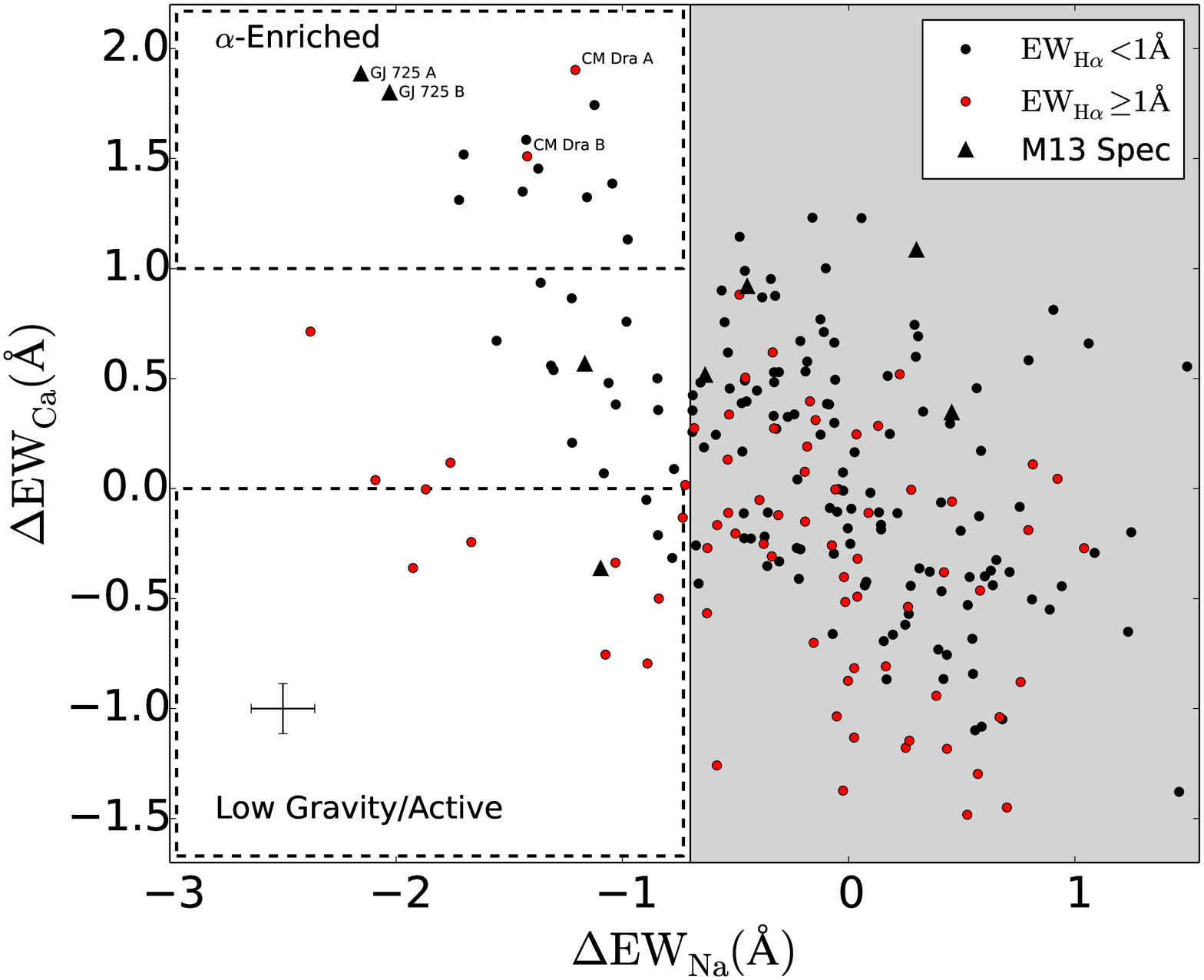}
\caption{\label{fig:cana}
The \dewca and \dewna values for M dwarfs in our sample, color-coded by their level of activity as indicated by H$\alpha$. We posit a simple interpretation: stars with low \dewna are likely either low-gravity and young, or $\alpha$-enriched and old, and \dewca can discriminate between these two due to filling-in related to chromospheric activity. Two M dwarf systems (CM Dra and GJ 725) with possible $\alpha$-enrichment \citep{Feiden:2014dn} are highlighted. Measurements on spectra from \citep{Mann:2013fv} are indicated separately and have been corrected for the use of a different instrument (0.96~\AA~offset). Mean errors for the stars observed in our program are indicated by the error bars in the lower left.
}
\end{center}
\end{figure*}

Figure~\ref{fig:cana} shows the plane defined by \dewna and \dewca. We note that, for the lowest \dewna values, targets with low and high H$\alpha$ are generally well-separated by \dewca (with the exception of CM Dra, an active close binary system). This is consistent with the behavior of \dewca being dominated by chromospheric activity, which is supported by our observation of \ca in emission and the findings of \citet{Kafka:2006kw}. We also note that CM Dra, a system that is suspected of $\alpha$-enrichment \citep{Feiden:2014dn}, stands out in this plane, as does the binary system GJ 725. In both systems, models with [$\alpha$/Fe]=0.2 provide a better match to observations than models with [$\alpha$/Fe]=0.0 (e.g. Figure \ref{fig:radlum}). We posit a simple interpretation of these features: stars with low \dewna are likely either low-gravity or $\alpha$-enriched, and the activity-sensitivity of \dewca can differentiate between these younger (more active) or older (less active) populations. As we only probe a limited range of [$\alpha$/Fe] and [Fe/H], this interpretation is not conclusive, but is strongly suggestive and provides a promising direction for further exploration and the development of better techniques for measuring M dwarf compositions.

\section{Conclusion}

The \ca and \naa spectroscopic indicators provide a wealth of new and easily obtained information about M dwarf stellar properties. The broad applicability of these physically-motivated indicators and associated techniques in mid- to late-type M dwarfs, their possible constraints on $\alpha$-enrichment, and their simplicity offer a substantial improvement over existing methods for measuring M dwarf parameters. These indicators and techniques will be useful for appraising M dwarf planetary systems discovered in the coming years (by e.g.~by \textit{TESS} and \textit{K2}), and for prioritizing nearby M dwarf exoplanet systems for planetary atmospheric characterization with \textit{JWST}. By providing stellar mass, luminosity, and radius through $M_K$, they will provide a necessary foothold for measuring planetary parameters and the location of the habitable zone.  They also provide a path toward more precise distances to Galactic M dwarfs, for which the current limiting factor is the scatter in the color-luminosity relations used to estimate distance \citep[e.g.][]{Bochanski:2013ky}. The Sloan Digital Sky Survey bandpass contains the \ca and \naa $I$-band features, as well as indicators of metallicity and \teff \citep{2011AJ....141...97W,2012AJ....143...67D}, potentially enabling archival analysis with the \naa and \ca indicators. Finally, we note that the \ca feature will be observed by the \textit{Gaia} RV Spectrometer and may provide leverage for the measurement of $\alpha$-enrichment and activity levels in the large sample of M dwarfs observed by \textit{Gaia}. The measurement of the features described here is relatively observationally inexpensive: a single $R\sim2000$ NIR spectrum is sufficient, and is substantially easier to obtain than the parallax measurements, high-resolution spectra, or broadband flux-calibrated spectra required by other techniques.

\begin{small}
	This work was partially supported by funding from the Center for Exoplanets and Habitable Worlds. The Center for Exoplanets and Habitable Worlds is supported by the Pennsylvania State University, the Eberly College of Science, and the Pennsylvania Space Grant Consortium. This work was also partially supported by the Penn State Astrobiology Research Center and the National Aeronautics and Space Administration (NASA) Astrobiology Institute (NNA09DA76A). We acknowledge support from NSF grants AST 1006676, AST 1126413, and AST 1310885. This work was partially based on data from the Infrared Telescope Facility. The Infrared Telescope Facility is operated by the University of Hawaii under Cooperative Agreement no. NNX-08AE38A with the National Aeronautics and Space Administration, Science Mission Directorate, Planetary Astronomy Program. 
\end{small}

\bibliographystyle{apj}

\clearpage

\end{document}